\documentclass[prd,twocolumn,superscriptaddress,preprintnumbers,nofootinbib]{revtex4-1}

\usepackage{amsmath}
\usepackage{amsfonts}
\usepackage{bbm}
\usepackage{microtype}
\usepackage{hyperref}
\usepackage{cleveref}
\usepackage{color}

\makeatletter
\DeclareRobustCommand{\rcite}[1]{%
  \rcite@aux#1,\@nil{#1}%
}
\def\rcite@aux#1,#2\@nil#3{%
  \if\relax#2\relax
    Ref.~\cite{#3}%
  \else
    Refs.~\cite{#3}%
  \fi
}
\makeatother

\begin{document}

\title{Uniformly expanding vacuum: \\a possible interpretation of the dark energy }

\author{Peng Huang}
\email{huangp46@mail.sysu.edu.cn}
\affiliation{School of Astronomy and Space Science, Sun Yat-Sen University, Guangzhou
510275, China}

\author{Fang-Fang Yuan}
\email{ffyuan@nankai.edu.cn}
\affiliation{School of Physics, Nankai University, Tianjin 300071, China}

\begin{abstract}

Following the spirit of the equivalence principle, we take a step further to recognize the free fall of the observer as a method to eliminate causes that would lead the perceived vacuum to change its original state. Thus, it is expected that the vacuum should be in a rigid Minkowski state or be uniformly expanding. By carefully investigating the impact on measurement caused by the expansion, we clarify the exact meaning of the uniformly expanding vacuum and find that 
this proposal may be able to explain the current observations of an accelerating universe.

\end{abstract}

\preprint{}

\maketitle
\section{Introduction}

The word vacuum in present work is defined to be the \textit{classical} spacetime perceived by a free fall observer, thus there is no gravity/gravitation in it. Usually the absence of gravity is treated as equivalent to the vanishing of Riemann tensor. However, in general cases, this is not so obligatory. In fact, as long as the observer is freely falling or geodesic (thus the effect of gravity is neutralized), the (local) spacetime is an vacuum perceived by this observer. For example, free fall in Minkowski spacetime is just inertial motion, and indeed corresponds to the vanishing of Riemann tensor; nevertheless, in de Sitter or anti-de Sitter spacetime free fall is comoving motion where the relevant Riemann tensor obviously does not vanish. A counter-example which deserves special attention is the Friedmann-Lema\^itre-Robertson-Walker (FLRW) universe. Although everything in this universe is freely falling, the perceived spacetime in radiation or matter dominated era is not an vacuum as defined here, which is because of the ubiquitous existence of perfect fluid. Furthermore, even if the dark energy dominates the universe, as long as it is a non-geometrically emergent effect of an unknown energy density, the perceived spacetime also cannot be looked as a vacuum defined here.

All in all, to define the vacuum in \textit{classical} level, free fall is a crucial element. From the perspective of the equivalence principle, free fall of the observer can remove the effect of gravity and thus results in an equivalent spacetime with no gravity which can be treated as the vacuum. Considering the general case with a cosmological constant whose value may equal to, larger or less than zero, the three possible vacua correspond to three solutions to the vacuum Einstein's field equation with the maximal symmetries. Thus, with the well-accepted perspective which insists that the symmetry is of central importance, it seems that these different vacua should have equal footing, therefore, one can expect that all of these different vacua will play important roles. Surprisedly, all the three vacua indeed have deeply penetrated into theoretical physics: Standard Model of particle physics is built on a Minkowski background \cite{Weinberg:1995mt}; the profound AdS/CFT correspondence \cite{Maldacena:1997re} endows the anti-de Sitter spacetime with special meaning.
Furthermore, it has been well accepted that the most economical way to explain the astonishingly accelerating expansion of our universe \cite{Perlmutter:1998np,Riess:1998cb} is to interpret the dark energy as a positive cosmological constant, the corresponding model has been named as "the concordance $\Lambda$CDM model".

However, while the $\Lambda$CDM model best fits the data, interpreting dark energy as a positive cosmological constant causes notorious difficulties \cite{Bousso:2007gp,Padilla:2015aaa}. The well known example is the serious fine-tuning problem related to it \cite{Weinberg:1988cp}. In addition, after years of efforts, a consistent quantum field theory in de Sitter background is still missing.
It is obvious that the combination of quantum field theory with cosmological constant has encountered many problems and subtle issues.
This in return implies that a whole new understanding of the vacuum is required to identify the nature of dark energy.

\section{Uniformly expanding vacuum}\label{s2}

\emph{What is the vacuum when there is no gravity}? We want to ask this question again but from some new angles and with new emphases. Following the spirit of the equivalence principle, we take a step further to recognize the free fall of the observer as a method to eliminate causes that would lead the vacuum to change its original state. The previously proposed question then turns into the following form: \emph{What state can the vacuum have when there are no causes responsible for its change}? Then, it is reasonable to expect that, in a well-defined context, the vacuum will maintain its original state if there is no cause to trigger its change. Taking this as a starting point, an inevitable answer to the above question is that
\emph{the vacuum should be in a rigid Minkowski state or be uniformly expanding}. However, while the rigid Minkowski state is interpreted as the metric of Minkowski spacetime does not evolve with time, the exact meaning of "uniformly expanding" is totally unclear at the moment, because the potential impact on measurement caused by the uniform expansion is yet to be clarified. Until such issue is settled down, no further definitive conclusions could be made.

To investigate the influence of uniform expansion of the vacuum on the measurement and to probe its true meaning, let us firstly imagine a Minkowski spacetime with a constant scale factor $a_1=1$ as
\begin{equation}
\label{2}
ds^2=-dt^2+1\cdot\big (dx^2+dy^2+dz^2\big ).
\end{equation}
Then, by kind of godlike ability, the original Minkowski spacetime is expanded into a new Minkowski spacetime with a new constant scale factor $a_2=2$, i.e.,
\begin{equation}
\label{3}
ds'^2=-dt'^2+4\cdot\big (dx^2+dy^2+dz^2\big ).
\end{equation}
Obviously the second metric is totally different from the original one. However, if the clock and ruler (it must be stressed that all the clock and ruler is spatially extended) in the spacetime expands synchronously, no change will be detected by these devices of measurement \cite{Hooft:2014daa}. This is equivalent to say that, despite being expanded twice as its original size, such a difference does not manifest on the level of the spacetime \textit{itself}.
The expansion of the spatial size in metric (\ref{3}) is only perceived by observers (us) with rigid \footnote{The word rigid here means that the measuring devices are bounded by their self-gravity or other gauge interaction and thus decouple from the expansion of the vacuum.} ruler. Furthermore, a direct deduction can be made by the observer (after figuring out the expansion of twice in spatial size) is that the time duration measured by the rigid clock is in fact only half of the time
that is pertinent to the spacetime structure \textit{itself}. Thus, the metric which the observer uses to describe the expanded Minkowski spacetime is
\begin{equation}
\label{4}
ds'^2=-4\cdot dt^2+4\cdot\big (dx^2+dy^2+dz^2\big ).
\end{equation}
In (\ref{2}) and (\ref{4}), time and spatial distance are measured by the \textit{same} rigid and space-extended devices, in particular, $t$ here \textit{always} denotes the \textit{proper time} perceived by the observer who operates these rigid measuring devices. The impact on the measurement of time caused by expanding twice as its original size is manifested by the prefactor $4$ before $dt^2$ in (\ref{4}). On the other hand, if one absorbs the prefactor $4$ into the four arguments and redefines them as new ones, then the originally straightforward interpretation of the measurement would become obscure.

The above warmup example is a special case in which the original Minkowski spacetime does not expand continuously but expand instantaneously into a new Minkowski spacetime. The rigid measurement made by us will inevitably lead the perceived spacetime to have the metric, before and after the expansion, denoted as (\ref{2}) and (\ref{4}) respectively. This result can easily be generalized into the continuous case as following: If the Minkowski spacetime expands with a time-dependent scale factor $a(t)$, and the measurements of space and time are implemented by space-extended and rigid devices, then the metric we will get to describe the new spacetime is
\begin{equation}
\label{5}
ds'^2=-a^2(t)dt^2+a^2(t)\big (dx^2+dy^2+dz^2\big ).
\end{equation}
Again, the spatial distance and time duration here have the same meanings as those in (\ref{4}), and $a(t)$ is measured after the measurements of space and time have been settled down. All these quantities are exactly those perceived by the observer with rigid and spatially extended measuring devices. The meaning of "uniformly expanding" is still unclear at present. Nevertheless, what one has known is that the metric describing the uniform expansion should has the form of (\ref{5}). The crucial point is that the proper time perceived by the observer is not the proper time pertinent to the vacuum \textit{itself}, and this difference is reflected in the prefactor $a^2(t)$ before $dt^2$ in (\ref{5}). If the observer continues to use his/her own proper time instead of the proper time of the uniformly expanding vacuum, the spacetime perceived is only a misleading illusion of the real one.

Now it is apparent that, if vacuum is indeed a conventional Minkowski spacetime, $a(t)\equiv1$; in contrast, if the vacuum has intrinsic uniform expansion, $a(t)$ can not be a constant. The question is how to deduce the form of $a(t)$ for uniformly expanding vacuum. One can make the pivotal step by noticing the fact that, the uniform expansion of the vacuum must be defined with respect to the proper time attributed to the vacuum, i.e., $a(t)dt$ must be rewritten as $dT$. Then it is straightforward to see that only when $a(t)=\exp (\lambda t)$, the metric
\begin{equation}
\label{0004}
  ds^2=-e^{2\lambda t}dt^2+e^{2\lambda t}(dx^2+dy^2+dz^2)
\end{equation}
can be arranged into
\begin{equation}
\label{004}
  ds^2=-dT^2+T^2(dx^2+dy^2+dz^2)
\end{equation}
with $T=\lambda^{-1}\exp (\lambda t)$, which qualifies to describe a uniformly expanding vacuum. This is how a uniformly expanding metric emerges finally. If observation confirms the existence of the exponential growth for the vacuum perceived by the free fall observer, the uniform expansion of the vacuum will be confirmed. However, for person who is unaware of the subtleties involved here, she/he would mistakenly use her/his own proper time instead of the proper time attributed to the vacuum, so that the metric (\ref{0004}) describing the uniformly expanding vacuum would be treated incorrectly as
\begin{equation}
\label{0005}
  ds^2=-dt^2+e^{2\lambda t}(dx^2+dy^2+dz^2),
\end{equation}
which means that the scale factor before $dt^2$ in (\ref{0004}) is ignored incorrectly. In this case, even if the exponential growth is observed, it will not be interpreted as the intrinsic uniformly expansion of the vacuum, but accelerating expansion caused by a cosmological-constant-like dark energy.

Let us say a little more words about the rewriting from (\ref{0004}) to (\ref{004}). A disturbing consequence of such a redefinition is: the new arguments will lose originally straightforward expressions of the measuring results, and thus may be misleading when one is not aware of subtle points here, for example, one may mistakenly regard the uniformly expanding vacuum simply as a Milne universe (see \cite{mkn} for some detail). Therefore, it needs to be stressed that, if the time is measured by a clock expanding synchronously with the expansion of the vacuum (corresponding to a redefinition), the structure of the vacuum shows itself as expanding uniformly with respect to a freely falling rigid ruler. On the contrary, under the condition that the time is measured by a freely falling, rigid and spatially extended clock, the uniformly expanding vacuum itself is seen to be expanding \textit{exponentially} with respect to a freely falling rigid ruler.

\section{Interpretation of the dark energy}

It is obvious that the value of the vacuum constant $\lambda$ must be very small otherwise there would be no existence of ourselves. A small $\lambda$ is hard to be detected by local experiments because the self-gravity of measurement devices may screen the effects caused by the uniform expansion of vacuum. Thus, it can be expected that any local experiment will show that the vacuum is Minkowskian. Only if the vacuum constant $\lambda$, if it exists, prevails over other energy components, such as the rigid measuring devices, then one has the possibility to precisely measure it. With these consideration, one can find that, for the purpose of determining whether the vacuum is Minkowskian or uniformly expanding, the ideal object one can observe is the recent (low redshift) FLRW universe. Such spacetime is flat, homogeneous and isotropic, furthermore, its average energy density of barotropic matter is smaller than almost any thing else (cosmological void in flat universe can be considered as a special exception). These characters make the recent FLRW universe a perfect mimic of the vacuum perceived by a free fall observer.

Embarrassingly, the concordance $\Lambda$CDM model tells that the vacuum is neither Minkowskian nor uniformly expanding, but is de Sitter which is described by following metric
\begin{equation}
\label{8}
ds^2=-dt^2+e^{2\Lambda t}\big (dx^2+dy^2+dz^2\big ).
\end{equation}
Before accepting such conclusion, we want to check carefully the processes that lead to (\ref{8}) at first. To start with, we notice the fact that (\ref{8}) is a special form of FLRW metric describing the universe, thus a close look at the FLRW metric may be helpful for our purpose. Generally, FLRW metric can be written as
\begin{equation}
\label{9}
ds^2=-dt^2+a^2(t)\big (\frac{dr^2}{1-kr^2}+r^2d\theta^2+r^2\sin^2\theta d\phi^2\big ),
\end{equation}
with $k>0,=0$ and $<0$ (not normalized) corresponding to spatially closed, flat and open universe, respectively. Then, measurements of spatial curvature $k$, cosmic time $t$ and spatial distance (equivalently, the scale factor $a(t)$) are implemented to determine its topology and evolution. Apparently, the entire evolution of this FLRW metric involves the entire history of the universe, and certainly will evolves different energy dominated periods. For our present purpose, the cosmological era which is of our central concern is the dark energy dominated era. The reason we focus on this special era is that, as mentioned earlier, this era mimics the vacuum better than any spacetime else in practice.

It has been shown in Sec.(\ref{s2}) that the intrinsic expansion (if exists) of the vacuum has nontrivial impacts on measurement of time. However, for measuring the spatial curvature,  detecting the value of $k$ is independent of the time and  the expansion of the spacetime does not change its topology, thus, one can trust the observation of the spatial curvature with $k=0$ \footnote{There is possibility for $k$ deviating from zero. However, it suffices for our purpose as long as $k$ is close enough to zero.} which means that the topology of the universe is flat. This is an important condition to ensure that one can safely treat the recent universe as a mimic of the vacuum perceived by the free fall observer.

For measuring the spatial distance and time, an important fact which is generally less emphasized is that any practical measurement of these quantities will inevitably involve rigid and spatial extended ruler. Nevertheless, the rigid ruler may involve in the measurement through an inconspicuous way. Taking the measurement of the luminosity distance of supernovae as an example. The luminosity distance is defined as
\begin{equation}
\label{10}
D_L=\sqrt \frac{L}{4\pi F}
\end{equation}
with $D_L$ the luminosity distance, $F$ the apparent luminosity denoting the energy received per second per square
centimeter of receiving area, $L$ the absolute luminosity denoting the energy emitted per second. Implementing this
definition into cosmology will involve factors related to cosmological redshift, however, (\ref{10}) is sufficient for illustrating
purpose. Then, at first glance, it seems that such luminosity distance is just the proportion between two different energy flux and
thus measuring it has no need to use rigid and spatially extended ruler. However, if one denotes $\mathcal{D}$ as radius of supernovae  and $\mathcal{F}$ the apparent luminosity on its surface, one has
\begin{equation}
4\pi D_L^2F=L=4\pi \mathcal{D}^2\mathcal{F},
\end{equation}
which follows by
\begin{equation}
D_L=\sqrt \frac{\mathcal{D}^2\mathcal{F}}{F}.
\end{equation}
Now it is obvious that, to measure the luminosity distance, rigid device is in fact necessarily needed. The apparent
calling for rigid device is to measure the radius of the supernovae, the hidden calling for rigid device is to measure
the proportion between $\mathcal{F}$ and $F$---the proportion can only be measured after a rigid square is given.

What is important is, the inescapability of the rigid and spatial extended devices in measuring time has crucial impact on the understanding of the current faint and exponential expansion of the universe. To see this, one can remind first that the cosmic time $t$ in metric (\ref{9}) is defined to be the proper time measured by a comoving observer \cite{mkn} or, in other words, the time measured in the rest frame of a comoving clock \cite{Weinberg:2008zzc}, which are equal to say that the cosmic time $t$ is idealized to be measured by a comoving clock in a single point. The key point here is that such hypothetical clock, localized on a single point, is the same as the clock expanding synchronously with the universe in the sense that they all can not feel and be impacted by the expansion of the spacetime. However, there is no such ideal clock in practice. On contrary, measurements that lead to the discovery of (\ref{9}) always involve measuring of time by spatially extended measuring devices. Furthermore, the self-gravity of the measuring devices always screen the local effects of the expansion of the spacetime.

Now comes the crucial point: in a universe which mimics the vacuum better than anything else, and with the measurements carried out \textit{not} with devices localized on a single point but rigid and spatial extended, the same logic which has been used in Sec.(\ref{s2}) that leads to metric (\ref{0004}) for the uniformly expanding vacuum must also be implemented here to ensure the correct interpretation of observations (see discussions related to (\ref{0005})). This is to say, the discovery of the faint and exponential expansion and attributing it to a cosmological constant like dark energy may in fact be interpreted as a confirmation of the speculation that the vacuum is neither Minkowskian nor in a de Sitter phase, but is uniformly expanding. Naturally, the observed $\Lambda$ is not the traditional cosmological constant or other mysterious dark energy but a vacuum constant $\lambda$ characterizing how fast the uniform expansion of the vacuum is.

\section{discussions}

Before any further investigation on the uniformly expanding vacuum, perhaps the first thing one must clarify is that the metric (\ref{0004}) is indeed a new metric describing an uniformly expanding vacuum but not be obtained from Minkowski or de Sitter spacetime by merely coordinate transformations. For this purpose, it suffices to observe that the Ricci scalar for metric (\ref{0004}) is
\begin{equation}
R_{UEV}=6\Lambda^2e^{-2\Lambda t},
\end{equation}
which is different from those for Minkowski (\ref{2}) or de Sitter (\ref{8}).
Nevertheless, it is worthy to take a more close look at this issue. For de Sitter spacetime, starting from (\ref{8}), one gets
\begin{equation}
\label{00011}
ds^2=-\frac{1}{\Lambda^2\eta^2}(d\eta^2-dx^2-dy^2-dz^2)
\end{equation}
with the conformal time $\eta=-\frac{1}{\Lambda e^{\Lambda t}}$. This metric has different scale factor with respect to metric for uniformly expanding vacuum, furthermore, $t$ in (\ref{0004}) stands for the proper time measured by an free falling observer with rigid and spatial extended devices, such explicit physical intension is lost in (\ref{00011}). Referring to Minkowski spacetime, let us focus on $D=2$ case for sharping the discussion. The uniformly expanding vacuum now is
\begin{equation}
\label{0009}
ds^2=-e^{2\lambda t}(dt^2-dx^2).
\end{equation}
On the other hand, the Minkowski spacetime in the proper coordinates ($t, x$), with $t$ the proper time and $x$ the distance measured by the uniformly accelerated observer, has an very similar form as following \cite{Mukhanov:2007zz}
\begin{eqnarray}
\label{00010}
ds^2&=&-(1+ax)^2dt^2+dx^2\nonumber \\
&=&-e^{2\tilde x}(dt^2-d\tilde x^2)
\end{eqnarray}
where $\tilde x=\frac{1}{a}\ln (1+ax)$.
From (\ref{0009}) and (\ref{00010}), one can see obvious difference between the uniformly expanding vacuum and Minkowski spacetime describing in an uniformly accelerating reference frame \footnote{See \cite{F. Rohrlich,E. A. Desloge,P. M. Brown,Munoz:2010pv} for detailed discussion on the relation between uniformly accelerated reference frame in flat spacetime and uniform gravitational field.}. This fact tells that these two concepts are really distinguishable. To sum up, all the above analysis shows that uniformly expanding vacuum is not a solution of vacuum GR equations with or without a cosmological constant, therefore the concept of uniformly expanding vacuum can be indeed considered as a new perspective on the accelerating expansion of the recent universe and the dark energy problem.

The question proposed in the present work is \emph{What state can the vacuum have when there is no cause leading its change}?
A similar question asked more than three hundred years ago is \textit {what motion can an object has when there is no external force acting upon it}? Newton's first law tells us that this object will either remain at rest or continue to move at a constant velocity. However, in responding to the question that with respect to what these states of motion are defined, Newton's first law must assume the existence of an absolute space which is not correct and has been abandoned. A natural suspicion about the concept of uniformly expanding vacuum then arises in a similar way: Whether the concept of uniformly expanding vacuum also needs absolute frame? The answer is no, since apparently the uniform expansion is defined with respect to free falling rigid and spatially extended measuring devices which have nothing to do with absolute frame. Furthermore, the tricky and also important point is, the uniform expansion of the vacuum has important impact on the measurement. Taking these factors into full consideration, one will inevitably find that the uniformly expanding vacuum is nontrivially realized as (\ref{0004}) with quantities in it has definite meanings according to the measurement. Therefore, one can say that the concept of uniformly expanding vacuum is not a trivial analogue of Newton's first law.

Due to the intrinsic and always exponential expansion of the uniformly expanding vacuum, a natural and unavoidable question is whether it is compatible with cosmological observations according to which the acceleration is only a recent phenomenon. Answering this question needs a complete theory of gravity whose vacuum solution is not Minkowski and (anti-) de Sitter spacetime but an uniformly expanding one, or all these different vacua are vacuum solutions of the underling theory, and the uniformly expanding vacuum is determined by observation \footnote{Conformal gravity is a potential candidate, nevertheless, since this theory is thought to be related to ghost issues, we leave this for further investigation.}. Only when such theory is given then one can figure out the evolution of FLRW metric in this theory. However, since such theory is not constructed yet, one is in fact incapable to answer the question proposed here.

Even though it is in this situation the following comment can still be made: The concept of uniformly expanding vacuum emerges from merely a careful reexamination of the equivalence principle and the impact on measurements caused by the expansion. Furthermore, the discovery of the faint and exponential expansion of the recent universe may be understood as a confirmation of this intrinsic character of the vacuum. Thus its validity is independent of concrete theory of gravity and can be treated as a fundamental principle on which the underling theory of gravity should base. In this perspective, the uniformly expanding vacuum implies (or, requires) an underling theory of gravity which should accommodate this new understanding of the vacuum and predict a correct history of the universe at the same time.

As a looking forward to the future work, the concept of uniformly expanding vacuum may have interesting influence on theoretical physics. For example, it will introduce a natural arrow of time due to the uniform expansion with respect to the rigid ruler, this also explains the appearance of the proper time $t$ in (\ref{0004}) (see \cite{Allahverdyan:2015qua} for an interesting discussion of the relation between the arrow of time and dark energy). Furthermore, because the current framework of quantum field theory is based on the special relativity, our new idea will necessarily lead to some modifications of its basic formalism including the Klein-Gordon equation and Dirac equation. Especially the quantum field theory in de Sitter spacetime needs to be reconsidered.

\begin{acknowledgments}
We would like to thank the anonymous referee for valuable and detailed suggestions which greatly improve the present work.      
We also thank Miao Li and Zhenhui Zhang for many inspiring discussions.
\end{acknowledgments}

\end{document}